\documentclass[preprint,showpacs,preprintnumbers,amsmath,amssymb]{revtex4}

\usepackage{graphicx}
\usepackage{dcolumn}
\usepackage{bm}
\usepackage{hyperref} 
\usepackage{latexsym}

\begin{document}

\title{Optimally  squeezed spin states}
\author{A. G. Rojo}\email{rojo@oakland.edu}
\affiliation{Department of Physics, Oakland  University, Rochester, Michigan
48309}

\begin{abstract}
  We consider  optimally spin-squeezed states that maximize the sensitivity of the Ramsey spectroscopy, and 
  for which the signal to noise ratio scales as the number of particles $N$. 
Using the variational principle we  prove   that  these states are eigensolutions 
of the  Hamiltonian 
$
H(\lambda)=\lambda S_z^2-S_x,
$ and that, for large $N$, the states become equivalent to the quadrature squeezed states of the harmonic oscillator.
We present numerical results that illustrate the validity of the equivalence.

\end{abstract}
\maketitle

\section{introduction}

Coherent control of ensembles  of two-level atoms
 may lead to the production of entangled spin-squeezed  states \cite{kitagawa}.
Each atom of the ensemble can be treated as a spin $1/2$, and, in analogy
with squeezed states of photons \cite{walls} and phonons \cite{phonons},
for  squeezed spin states  the quantum noise
  in the different components of the total spin is redistributed preserving the uncertainty relation. 
 Depending on the context, different definitions of spin squeezing can be used. 
Starting from the uncertainty relation $\Delta S_x \Delta S_y \ge |\langle S_z\rangle/2|$
(and cyclic permutations),
a possibility is to define  states satisfying $\Delta ^2 S_i < 
 |\langle S_j\rangle/2|$
as
spin-squeezed\cite{eberly}.   
Another definition was proposed by Wineland {\em et al.} \cite{wineland}, who 
showed that the resolution in spectroscopic experiments on $N$ two-level atoms is 
determined by the factor

\begin{equation}
\xi=\frac{\Delta S_{\perp} }{ |\langle {\bf S}\rangle|},
\label{xiwineland}
\end{equation}
which measures the quantum noise in a direction perpendicular to the mean value
 of the total spin.
States with small $\xi$  are relevant to quantum
 information and constitute the focus of our paper. We will also discuss Kitagawa and Ueda's\cite{kitagawa} squeezing parameter given by 
  
\begin{equation}
\xi_{\rm spin}=\frac{\Delta S_{\perp} }{{ |\langle {\bf S}\rangle|}^{1/2}},
\label{xispin}
\end{equation}
as well as the parameter $\xi_z$ proposed  by Raghavan {\it et al.}\cite{raghavan}:

\begin{equation}
\xi_{z}=\frac{\Delta S_{\perp}^2 }{ S^2-\langle { S_z}\rangle^{2}}.
\label{xiz}
\end{equation}

From the commutation relation for the spin components in three orthogonal directions we have that
\begin{equation}
{\Delta S_z\over \langle S_x\rangle}\Delta S_y\ge {1\over2}.
\end{equation}

Since the fluctuations for $N$ spins $1/2$ are
 bounded [$\Delta {\bf{S}}\le \sqrt{N/2(N/2+1)}$], there is a 
lower bound (the  Heisenberg limit) for  
the squeezing factor:
\begin{equation}
\xi >{1\over N}
\end{equation}
for large $N$. 
On the other hand, for product states (or unentangled states) $\xi = 1/\sqrt{N}$. The simplest example is the 
completely polarized state in the $x$ direction $|\rightarrow, \rightarrow, 
\rightarrow, \cdots, \rightarrow \rangle$, which has    
$\langle S_x\rangle=N/2$, $\langle S_y\rangle=\langle S_z\rangle=0$, 
and $\langle \Delta S_z\rangle=\langle \Delta S_y\rangle=\sqrt{N}/2$.

Given the above bounds, the key to approaching the Heisenberg limit is to find states with $|\langle S_x\rangle|$ of order $N$ and a
fluctuation $\Delta S_z$  of order one. 

Consider for example $N=3$ with $S=3/2$. 
The Hilbert space has dimension 4, and we can label the states by their projection $m=-3/2,-1/2,1/2,3/2$
 along the $z$ direction.
 The simplest squeezed state $|\Psi_{S=3/2} \rangle$ has the form \cite{yurke}
\begin{equation}
|\Psi_{S=3/2} \rangle={|1/2\rangle+|-1/2\rangle\over\sqrt{2}}.
\label{simple}
\end{equation}

This state is polarized in the $x$ direction, with $\langle\Psi_{3/2} |S_x|\Psi_{3/2} \rangle=1$ ,
$\langle\Psi_{3/2} |S_y|\Psi_{3/2} \rangle=\langle\Psi_{3/2} |S_z|\Psi_{3/2} \rangle=0$ and 
 $\langle\Psi_{3/2} |S_z^2|\Psi_{3/2} \rangle=1/4$, giving $\xi=1/2<1/\sqrt{3}$.  
  
If we include the states with maximum projection ($S_z=\pm3/2$) we will increase both $\Delta S_z$ and $\langle S_x\rangle$ and the question
is whether the inclusion of these states can improve the ratio $\Delta S_z/\langle S_x\rangle$. The
 natural step is to consider a state of the form:

\begin{equation}
|\Psi'_{S=3/2} \rangle={1\over\sqrt{1+a ^2}} \left[ {|1/2\rangle+|-1/2\rangle\over \sqrt{2}}
+a {|3/2\rangle+|-3/2\rangle\over \sqrt{2}}\right],
\end{equation}
for which 

\begin{equation}
\xi={1\over 2} {\sqrt{(1+a^2)[1+(3a)^2]}\over 1+\sqrt{3}a}\simeq {1\over 2}(1-\sqrt{3}a), \;\;\;\;\;(a \ll 1).
\end{equation}

This result indicates that  the inclusion
 of larger spin projections with a small amplitude  can improve 
the spin squeezing. For $a =1/\sqrt{3}$ the state has well defined
 $S_x=3/2$, and $\xi=1/\sqrt{3}$. For $a \simeq 0.16 $  the squeezing factor has 
its minimum value $\xi \simeq 0.44$.

In section \ref{secv}  we  prove, using the variational principle,  that the states that minimize $\xi $ (the optimally squeezed states) are solutions 
of the  Hamiltonian 
\begin{equation}
H(\lambda)=\lambda S_z^2-S_x,
\label{hmu}
\end{equation}
where, without loss of generality, we
 chose the $x$ axis in the direction of the mean spin and $S_{\perp} =S_z$. 

 For different values of the parameter $\lambda$, the 
 eigenstates of (\ref{hmu}) have  the minimum $\Delta S_z$
 for a given $\langle S_x \rangle$ and, in the large $N$ 
limit, the squeezing factor
  scales  as $\xi \sim 1/N$.  For $1\ll \Delta S_z \ll \sqrt{N}$,  
$H(\lambda)$ can be mapped to the harmonic oscillator Hamiltonian, and the optimally 
squeezed states become
  gaussians of reduced variance, identical to the quadrature squeezed states.
In section \ref{secn} we  present numerical diagonalizations of (\ref{hmu}) for up to 600 spins to illustrate the 
validity of the mapping. 

\section { { Variational calculation}}
\label{secv}

For $N$ spins, the  general state with $\langle {\bf S}\rangle$  in the $x$ direction can be written  
 as the following superposition of Dicke states $|S,m\rangle$ (eigenstates of the total spin $S$ and 
the spin projection $m$ in the $z$ direction):

\begin{equation}
|\Psi\rangle = \sum_{m,S} \varphi_{S,m} |S,m\rangle,
\end{equation}
with $\varphi_{S,m}$ real to guarantee that $\langle\Psi|S_y|\Psi\rangle=0$.
The squeezing factor can be written as  the following functional of the set $\{ \varphi_{S,m} \}$:
\begin{equation}
\xi (\{\varphi_{S,m}\}) = { \left[ {\sum_{S,m} m^2\varphi_{S,m}^2}\right]^{1/2}  \left[ \sum_{S,m}\varphi_{S,m}^2 \right]^{1/2} \over  \sum_{S,m} 
{\varphi_{S,m}}X_{S,m}\varphi_{S,m+1} 
} ,
\label{xivar}
\end{equation}  
with $m$ integer (half integer) for $N$  even (odd), and \begin{equation}X_{S,m}=\sqrt{(S+m+1)(S-m)}
\end{equation}
 twice the matrix element of
 the operator $S_x$.
The minimization  condition 
\begin{equation}{\delta \xi \over \delta \varphi _{S,m} }
=0\end{equation} 
gives rise to the following difference equation for $\varphi_{S,m}$  
\begin{equation}
\lambda  m^2 \varphi_{S,m} -
{1\over 2}
\left(X_{S,m}{\varphi_{S,m+1}} +   
X_{S,m-1}{\varphi_{S,m-1}}
\right)
=E(\lambda) \varphi_{S,m} ,
\label{diff}
\end{equation} 
with $\lambda$ and $E(\lambda)$ functions of the sums in Eq.(\ref{xivar}).  Eq.(\ref{diff}) is identical to the eigenvalue equation for 
the Hamiltonian 
$
H(\lambda)=\lambda S^2_z-S_x$,
 and in order to find the  minimum squeezing 
 we can regard $\lambda$ as a parameter and  minimize $\xi $  with respect to $\lambda$ for each eigenstate of $H(\lambda)$. Also, 
since $H(\lambda)$ is even in the projection in the $z$ direction,
the solutions are even and 
odd in $m$, and  $\langle\Psi|S_z|\Psi\rangle=0$. 
It is interesting to note that Eq.(\ref{diff}) corresponds 
to a one dimensional tight-binding Hamiltonian for a ``particle"
in a parabolic potential of strength $\lambda$  and hopping to near neighbors given by 
$X_{S,m}$.  
 The squeezing factor can then be interpreted as the ratio of a mean 
square displacement and a mean kinetic term.  
For large $S$ the parabolic potential confines the particle to small values of
$m$ and as a result $\langle m^2\rangle \sim O (1)$,  and 
 $\langle S_x\rangle \sim O (S)$, 
which implies that $\xi$ is minimized by states
 corresponding by the maximum total spin. We work in this subspace and in what follows we omit the subindex $S$ in $\varphi_{S,m}$.

%
%
%
%
Hamiltonian (\ref{hmu}) 
 was considered previously by Law {\it et al.} to generate spin-squeezed states \cite{law}, and also 
arised several times in the context of Bose-Einstein condensation.  Specifically,  if we write the spin operators in term of the  Schwinger bosons
$a^{\dagger}$ and $b^{\dagger}$:
\begin{equation}
S_z={1\over 2} \left(a^{\dagger}a -b^{\dagger}b\right), \;\;\;\; S^{+}=a^{\dagger}b, 
\end{equation} 
Hamiltonian   (\ref{hmu}) becomes (ommiting terms that depend on the total number $N$)
\begin{equation}
H(\lambda)=2\lambda \left[ \left( a^{\dagger} a\right)^2 +\left( b^{\dagger} b\right)^2\right]-{1\over 2} \left(a^{\dagger}b+b^{\dagger}a \right). 
\end{equation}
   
The Hamiltonian above corresponds two condensates, each of them treated in the single mode approximation, with the $S_x$ term in (\ref{hmu}) 
becoming a Josephson coupling betwen the condensates. This correspondence was used by Milburn {\it et al.}\cite{milburn}, Steel and Collett\cite{steel},  
Gordon and Savage \cite{gordon} and Raghavan {\it et al.}\cite{raghavan}. 
In this representation, the squeezing parameter $\xi$ can be interpreted
as the ratio of the particle number fluctuation ($\Delta S_z$) to the Josephson energy ($\langle S_x \rangle$) of two weakly coupled condensates: 
 start 
with two condensates ($a$ and $b$) with equal particle number $N_a=N_b=N/2$. This corresponds to 
the Dicke state $|N/2,0\rangle$ with $S_z=0$. Now we turn on a small Josephson coupling between the condensates that will
induce a small fluctuation in the particle number of each condensate $\Delta (N_a-N_b) \equiv \Delta S_z\sim {\cal O}(1)$. On the other hand, 
for the  Josephson energy  we have 
\begin{equation}
\langle ab^{\dagger}\rangle \simeq  \langle a\rangle\langle b^{\dagger}\rangle \sim {\cal O}(N),
\end{equation} 
giving $\xi \sim 1/N$.

%

Before analyzing the detailed form of the solutions of (\ref{diff}) we  can gain   insight on the optimally squeezed states by examining the 
general behavior of $E(\lambda)$ for small and large $\lambda $, 
and using the Hellman-Feynman theorem\cite{feyn}
\begin{equation}
{d E\over d \lambda}=\left \langle {\partial H\over \partial \lambda} \right\rangle
\end{equation} 
to compute
 the averages: 
 $\langle m^2\rangle ={dE(\lambda)/ d\lambda}\equiv (\Delta S_z)^2,
$
and 
 $
\sum_m \varphi_mX_{S,m} \varphi_{m+1}
=\lambda {dE(\lambda)/ d\lambda}-E(\lambda)\equiv \langle S_x\rangle$, giving
\begin{equation}
\xi(\lambda)={\left[{dE(\lambda)\over  d\lambda}\right]^{1/2}\over \lambda {dE(\lambda)\over  d\lambda}-E(\lambda)}.
\end{equation}
For  small $\langle m^2\rangle $ (large
 $\lambda $), 
 the solutions for 
even and odd $N$ are qualitatively different.
  For $\lambda \rightarrow \infty$ and $N$ 
even, the ground state  
energy
 the solution corresponding to the ``particle"  
at the bottom of the potential well ($m=0$).
Using perturbation  theory keeping the lowest lying states
 ($m=0,\pm 1$), 
\begin{equation} 
E(\lambda) ={\lambda\over 2}
-\sqrt{ \left({\lambda\over 2}\right)^2+{N^2\over 8}}
\simeq -{N^2\over 8 \lambda},
\end{equation} 
for $N$ even.
On the other hand, for odd $N$  the 
particle can be either at $m=+1/2$ or $m=-1/2$, the energy diverges as 
$\lambda/4$, and the ground state corresponds to a linear combination as in Eq.(\ref{simple}).
The condition $d\xi /d \lambda =0$ 
implies 
\begin{equation}
{d^2E\over d
\lambda^2}{d\over d\lambda}[\lambda E(\lambda)]=0.
\end{equation} 

Since $E(\lambda)$ has monotonic curvature, the minimum is given by the product
$\lambda E(\lambda)$. Given the discussion above,  for
$N$ even the minimum value of $\xi $ corresponds to $\lambda \rightarrow \infty$:
\begin{equation}
\xi^{(\rm min)} ={ \sqrt{2}\over N}, \;\;\;\;\;\; ({\rm even}\, N).
\end{equation}

This absolute minimum is however not very interesting since it corresponds 
to a ratio of zeroes therefore implying a vanishing mean value of the spin. This 
limit has been  pointed out by Andr\'e and Lukin \cite{andre} for a perturbative
 calculation. For odd $N$, since $E(\lambda)$ diverges for large $\lambda$, the product $\lambda E(\lambda)$ 
changes sign and  the absolute minimum of $\xi $ corresponds to a large
 but finite value of $\lambda$ that we determine numerically
 by solving exactly Hamiltonian (\ref{diff}). 

If we extend our variational treatment to the squeezing parameter given by Eq.(\ref{xispin}), we see that
the states that minimize $\xi_{\rm spin}$ are also eigenstates of (\ref{hmu}), but this time we have to minimize 
$\xi_{\rm spin}(\lambda) $ given by

\begin{equation}
\xi_{\rm spin}(\lambda)={\left[{dE(\lambda)\over  d\lambda}\right]^{1/2}\over \left[\lambda {dE(\lambda)\over  d\lambda}-E(\lambda)\right]^{1/2}}.
\label{xispinlam}
\end{equation}

The minimum value of $\xi_{\rm spin}$ also corresponds to $\lambda \rightarrow \infty$, but the square root in the denominator of
(\ref{xispinlam}) implies that

\begin{equation}
\xi_{\rm spin}^{({\rm min})}=0.
\label{xispinmin}
\end{equation}

On the other hand, the variational calculation shows that the states that minimize $\xi_z$ given by Eq.(\ref{xiz}) are eigenstates of 
\begin{equation}
H_{\rm s}(\lambda)=\lambda S_z^2-S_z.
\end{equation}

As discussed in Ref. \cite{raghavan}, the states that minimize $\xi_z$ are eigenstates of $S_z$. 

\section{Numerical results}
\label{secn}
\begin{figure} 
 \vspace*{0.cm}
 \includegraphics*[width=0.45\textwidth]{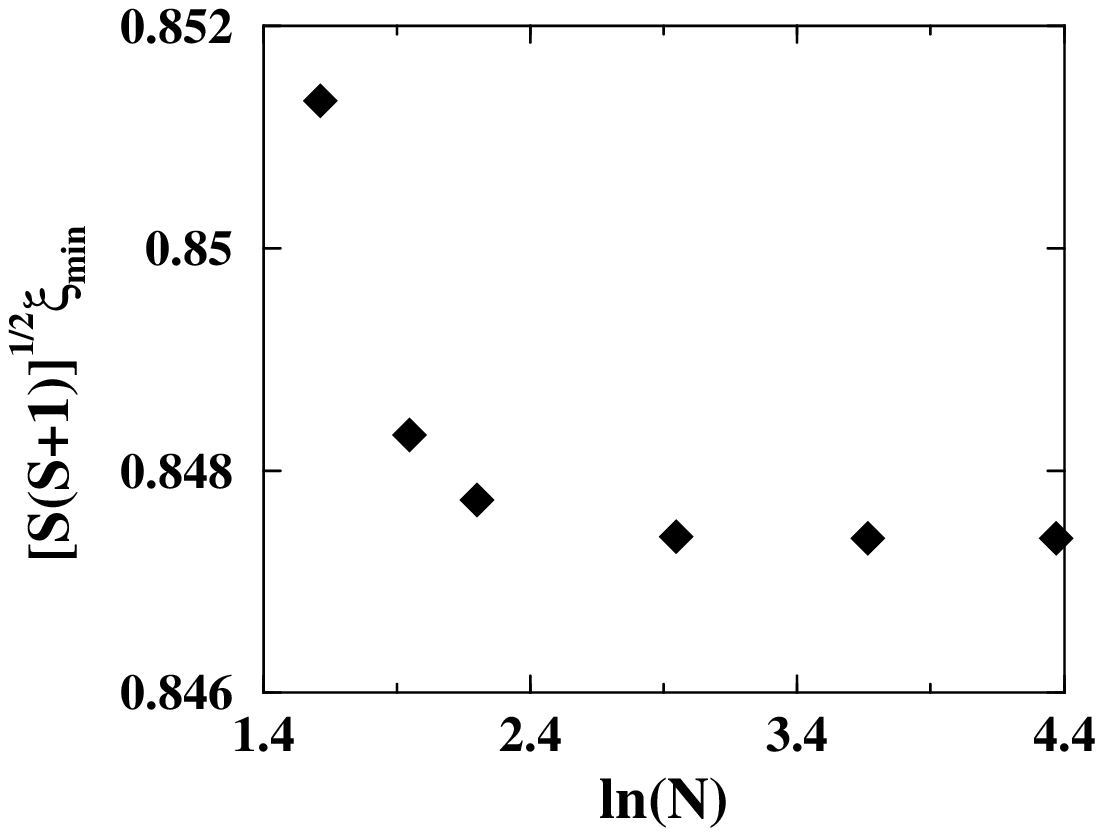}
\caption{Numerical results of the absolute minimum $\xi _{\rm min}$ for even $N$ (and $S=N/2$) as a function of $N$. }
\label{figure0}
\end{figure}

The numerical  minimization for different $N$ (see  Figure \ref{figure0})
 shows that, in  the large $N$ limit:

\begin{equation}
\xi_{\rm min} ={ 1.695\over N}, \;\;\;\;\;\; ({\rm odd}\, N).
\end{equation}

In general, for larger values of $\xi$    
 the solutions of (\ref{diff}) give  the  optimally 
squeezed states  with minimum $\Delta S_z$ for a given value of $\langle 
S_x\rangle$, and  for which $\xi $ scales as $1/N$. 
In order to show this, we first 
note that the {\em exact}
 normalized eigenstate of $S_x$  with eigenvalue $N/2$ and $N$ even
 \cite{oddn} has the form of a gaussian of width $\sqrt {N}:$

\begin{equation}
\varphi_m^{(N/2)}=2^{-N/2}\binom{N}{{N\over 2}-m}\simeq \left({2\over \pi N}\right)^{1/4} e^{-m^2/N}.
\end{equation}

The eigenstate corresponding to $S_x=-N/2$ is simply $ \varphi_m^{(-N/2)}=(-1)^m\varphi_m^{(N/2)}$. 
Using the explicit expressions for $X_{N/2,m}$ it is straightforward to show that the states with smaller projections $S_x=N/2-n$ are of form 
$ \varphi_m^{(N/2-n)}\propto h^{(n)}_{N\over 2}(m)\varphi_m^{(N/2)}$, and   $h^{(n)}_{S}(m)$, with $S=N/2$, obeying the following recursion relation:
\begin{equation}
{N\over4}\left[
h^{(n)}_{S}(m+1)
+h^{(n)}_{S}(m-1)
\right]
-
{m\over2}\left[
h^{(n)}_{S}(m+1)-
h^{(n)}_{S}(m-1)
\right]
=\left({N\over 2}-n\right)
h^{(n)}_{S}(m)
.
\label{hermdisc}
\end{equation}

For large $N$  we can take the continuum limit of $h^{(n)}_{S}(m)$ and (\ref{hermdisc}) becomes the differential equation satisfied
by the Hermite polynomials $H_n(x)$ \cite{arfken}, with $x=m\sqrt{2/N}$. 
This proves that, in the large $N$ limit, the eigenstates $ \varphi_m^{(N/2-n)}$ of $S_x$ in the $S_z$ representation are harmonic oscillator wave functions
$\varphi_n(x)$, with the following operator equivalences:
\begin{eqnarray}
S_z &\rightarrow& \sqrt{N\over 2}x, \nonumber \\
S_y &\rightarrow& i\sqrt{N\over 2}{d\over dx} \nonumber \\
S_x &\rightarrow& -{N+1\over 2}+{1\over 2}\left(-{d^2\over dx^2} +x^2\right),
\label{map1}
\end{eqnarray}
and an implicit truncation in the allowed excitations $n$ of the harmonic oscillator Hilbert space: $n<N/2$. 
[For quantum numbers $N/2<n<N$, the solutions are of the form $\varphi_m^{(N/2-n)}= (-1)^m \varphi_{N-n}(m\sqrt{2/N})$.]
The operator equivalence of Eq. (\ref{map1}) corresponds to the well known Holstein-Primakov transformation\cite{kittel} provided $\Delta S_x/N \ll1$. 
Also, the correspondence between the  Harmonic oscillator algebra and that of the angular momentum  was discussed using the so-called group contraction
by Arecchi {\it et al.} \cite{arecchi}. 
In this formulation, (\ref{hmu}) becomes the harmonic oscillator Hamiltonian and the ground state is a rescaled gaussian
 $\varphi_0(x)={\pi}^{-1/4} {\sigma}^{-1/2}\exp{-x^2/2\sigma^2}$, which, in the discrete representation has the form

\begin{equation}
\varphi_m \simeq { 1\over (2 \pi)^{1/4}  \Delta S_z^{1/2}}e^{-(m/ 2 \Delta S_z)^2},
\label{solution1}
\end{equation}
with
$\Delta S_z= \sigma \sqrt{N}/2$. Since the  mapping to the continuum is exact 
provided
$1\ll \Delta S_z \ll N$, taking the limit of large $N$ with $\Delta S_z$ fixed we obtain 

\begin{equation}
|\langle S_x \rangle| = {N+1\over 2} -{1\over 4} \left (\sigma^2+{1\over \sigma ^2} \right)\simeq{
N\over 2} \left [ 1- {1\over 8(\Delta S_z)^2 }\right].
\end{equation}

\begin{figure}
\includegraphics*[width=0.4\textwidth]{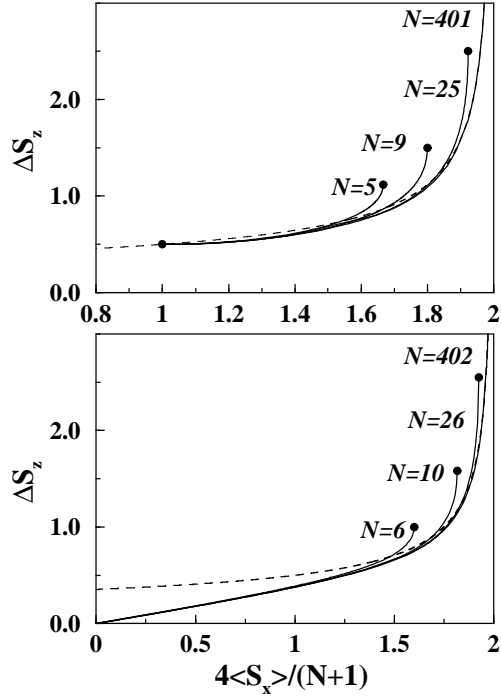}
\caption{$\Delta S_z$ vs $\langle S_x \rangle $ for the optimally squeezed states for even and odd number of spins $N$ obtained by numerical diagonalization of Eq.(\ref{diff}). 
The end points for $N$ even and odd correspond to the
 eigenstates of $S_x$  (with eigenvalue $N/2$) and  $\Delta S_z=\sqrt{N}/2$. For odd $N$ there is a minimum value of $\langle S_x \rangle =(N+1)/4$ 
that corresponds to $\Delta S_z=0.5$. This is the bifurcation identified by S\o rensen \cite{sorensen} for $N=3$ and discussed in the text.   
The dashed lines correspond to the analytical result 
$\Delta S_z=(1/\sqrt{8})/\sqrt{1-2\langle S_x\rangle/N }$.}
\label{evenodd}
\end{figure}

\begin{figure}
\includegraphics*[width=0.4\textwidth]{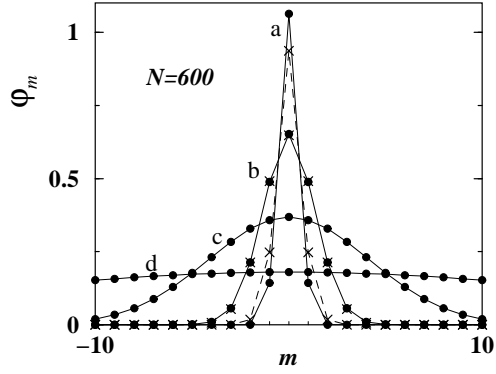}
\caption{
Comparison between the exact eigenstates of $H=\lambda S_z^2 - S_x$ for $600$ spins
(circles)  and the solutions of the 
discretized harmonic oscillator of Eq.(\ref{solution1}) ($\times$'s) in the continuum approximation.
The curves correspond 
to $\Delta S_z/ \Delta S_{z,0}=0.03$ (a), 0.08 (b), 0.24 (c) and 1.0 (d),
with $\Delta S_{z,0}=\sqrt{N}/2$. For 
$\Delta S_z> 0.3 \Delta S_{z,0}$ the two solutions are indistinguishable.}     
\label{figure2}
\end{figure}

In Figure \ref{evenodd} we present comparisons of the  continuum approximation with exact diagonalizations of (\ref{diff}) that show excellent agreement for large $N$ 
and $\Delta S_z$ as small as $0.5$. In Figure \ref{figure2} we show comparisons of the exact eigenstates $\varphi_m$ of Eq. (\ref{diff}) with the analytical solution.

Notice that the operator $S_y$ in the  discrete (or tight-binding) representation has the form of a 
velocity operator: 
$S_y=(i/2)\sum_m(X_{N/2,m}|m+1\rangle \langle m| - {\rm h.c.})$. In the  tight-binding limit the commutator of  
the  position (corresponding to $S_z$) and the velocity operator is not a 
  constant but is proportional to the 
kinetic energy operator (corresponding to $S_x$) and the commutation relations 
are equivalent. In the continuum approximation $S_y \rightarrow i\sqrt{N\over 2}{d\over dx}$ and 
 $\Delta S_y$  is given by

\begin{equation}
\Delta S_y =
\left\langle- {N\over 2}{d^2\over dx^2}\right \rangle^{1/2}
 ={N\over 4\Delta S_z} ,
\end{equation} 
which proves that  the optimally squeezed states in the large $N$ limit (and for 
$1\ll \Delta S_z \ll N$) correspond to minimum uncertainty wave-packets satisfying 
$\Delta S_z \Delta S_y = \langle S_x\rangle /2$.
Note that the optimally squeezed states given by Eq. (\ref{solution1}) can be written in the form
\begin{equation}
\Psi(N,M,\alpha)=C \exp{(-\alpha S_z^2)} \exp{(-i\pi S_y/2)}|N/2,M\rangle,
\label{norash}
\end{equation}
with $\alpha=1/(2\Delta S_z)^2-1/N\simeq 1/(2\Delta S_z)^2$ and $M=N/2$. The state (\ref{norash}) has some similarity
with the minimum uncertainty states considered by Rashid \cite{rashid,puri}, but the $S_z^2$ (as opposed to  $S_z$ for the Rashid states) 
in the first exponential guarantees that
the squeezing is in the direction perpendicular to the mean spin for all $M$. 

\begin{figure}
\includegraphics*[width=0.4\textwidth]{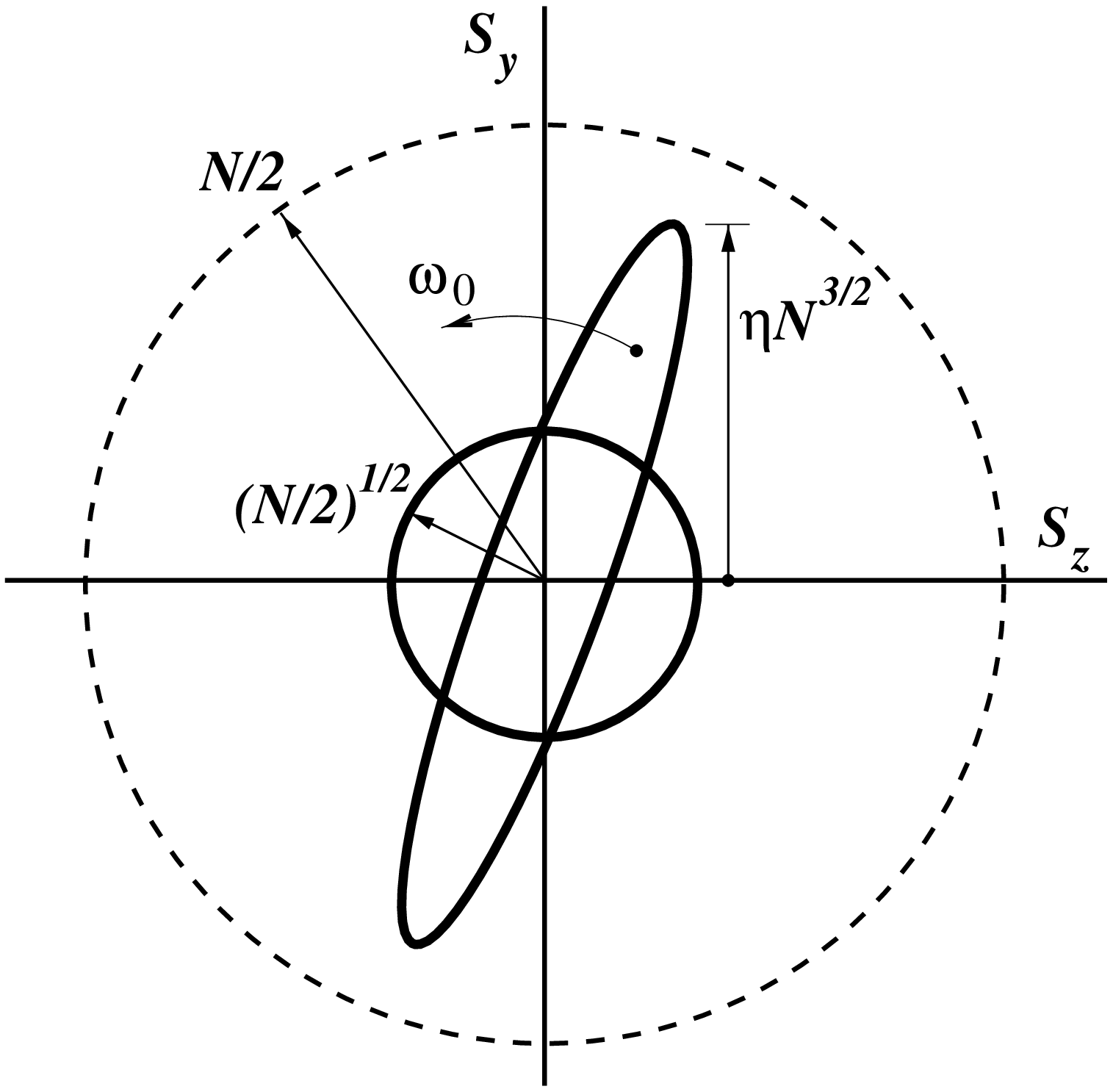}
\caption{
Quasiprobability distribution in the $(S_z,S_y)$ plane  for $N$ spins, before and after a pulse $H'=
\delta (t) \eta S_z^2$ is applied on the lowest  eigenstate of $H_0=\omega_0 S_x$. In contrast with the harmonic oscillator, the distribution is bounded by a circle
of radius $N/2$.}
\label{figure4}
\end{figure}

We close by noting that the harmonic  oscillator equivalence extends to the dynamic generation of squeezed states by the application of an impulsive non-linear Hamiltonian. Consider an ensemble of
identical two-level atoms with energy splitting $\hbar \omega_0$. We define the corresponding spin quantization axis in the $x$ direction so that $H_0=\omega_0S_x$,
and the equations of motion for $S_z(t)$ and $S_y(t)$ are exactly the same as those of $x$ and $p$ for a harmonic oscillator of frequency $\omega_0$.
If a pulse of the form $H'=
\delta (t) \eta S_z^2$ is applied, the quasiprobability distribution in the $(S_y,S_x)$ plane is modified as in Figure \ref{figure4}, the only difference with the harmonic oscilator
being that  $\eta<1/\sqrt{N}$ (for larger $\eta$ the response is periodic in $\eta$).  For times $t>0$ the distorted distribution rotates at frequency $\omega_0$ and the squeezing factor evolves as  
\begin{equation}
\xi(t)=\frac{1}{\sqrt{N}} \frac{\left[1+(\eta N \sin \omega_0t)^2 -\eta N \sin 2\omega_0t\right]^{1/2}}{1-{N\eta^2/2}}.
\end{equation}

If we call $\eta =\alpha_0/\sqrt{N}$, with $\alpha_0<1$, the minimum squeezing $\xi_{\rm min}=([1-\alpha_0^2/2]\alpha_0N)^{-1} $  scales as $1/N$ and is reached 
twice during the cycle of the rotation of the ellipse of Figure \ref{figure4}.

\acknowledgments
We thank Paul Berman, Tony Bloch and Ken Elder for very valuable conversations. I am indebted to Alejandro Uribe for referring me to the Krawtchowk polynomials.   

\newpage
\appendix

{\bf Appendix}

In this section we will derive the recurrence equation for the discrete  polynomials of Eq. (\ref{hermdisc}).  The normalized eigenstate $|n\rangle_x$ of  $S_x$ corresponding to the eigenvalue $S-n$, with $S=N/2$,  can be written as 

\begin{equation}
|n\rangle_x = \sum _ {m=-S}^S \varphi_S^{(n)}(m)|m\rangle_z,
\label{nx}
\end{equation}
with $|m\rangle_z$ the eigenstate of $S_z$ with eigenvalue $m$. For our choice of quantization axes we have 
\begin{eqnarray}
S_z|n\rangle_x& =&\frac{\left(S^++S^-\right)}{2}|n\rangle_x \\
&=&\frac{\sqrt{2S}}{2}\left[  \left(1-\frac{n-1}{2S}\right)^{1/2}\sqrt{n}|n-1\rangle_x +
\left(1-\frac{n}{2S}\right)^{1/2}\sqrt{n+1}|n+1\rangle_x
\right],
\end{eqnarray}
which, substituted in (\ref{nx}) gives
\begin{equation}
2\left(\frac{m}{\sqrt{S}}\right)\varphi_S^{(n)}(m)=\left(1-\frac{n-1}{2S}\right)^{1/2}\sqrt{2n}\varphi_S^{(n-1)}(m)
+\left(1-\frac{n}{2S}\right)^{1/2}\sqrt{2(n+1)}\varphi_S^{(n+1)}(m).
\label{rec1}
\end{equation}

Now we define $h^{(n)}_S(m)$ as
\begin{equation}
\varphi_S^{(n)}(m)=\frac{h^{(n)}_S(m)}{\sqrt{2^nn!}}2^{-S}\binom{2S}{{S}-m},
\end{equation}
and subtitute in (\ref{rec1}) to obtain the following recursion relation
\begin{equation}
h^{(n+1)}_S(m)=2\left(\frac{m}{\sqrt{S}}\right)
 \frac{1}{\sqrt{1-{n}/{2S}}} h^{(n)}_S(m)-2n\sqrt{\frac{1-(n-1)/2S}{1-n/2S}} h^{(n-1)}_S(m)
\label{rec2}
\end{equation}

Note that, in the limit $S\rightarrow \infty$, $m/\sqrt{S} \rightarrow x$, $h^{(n)}_S(m)\rightarrow H_n(x) $, and (\ref{rec2}) becomes
the recurrence equation for the Hermite polynomials:
\begin{equation}
H_{n+1}(x)=2xH_n(x) -2nH_{n-1}(x).
\end{equation}

Iterating (\ref{rec2}) we obtain the first few discrete polynomials $h^{(n)}_S(m)$:

\begin{eqnarray}
h^{(0)}_S(m)&=&1\nonumber \\
h^{(1)}_S(m)&=&2\frac{m}{\sqrt{S}}\nonumber \\
h^{(2)}_S(m)&=&\frac{4\left(\frac{m}{\sqrt{S}}\right)^2-2}{\sqrt{1-1/2S}}\nonumber \\
h^{(3)}_S(m)&=&\frac{8\left(\frac{m}{\sqrt{S}}\right)^3-4\left(\frac{m}{\sqrt{S}}\right)(3-1/S)}{\sqrt{(1-1/2S)(1-1/S)}}\nonumber \\
h^{(4)}_S(m)&=&\frac{16\left(\frac{m}{\sqrt{S}}\right)^4-48\left(\frac{m}{\sqrt{S}}\right)^2(1-2/3S)+12(1-1/S)}{\sqrt{(1-1/2S)(1-1/S)(1-3/2S)}}.
\end{eqnarray}

Finally we note that the above discrete Hermite polynomials are proportional to the Karwtchouk polynomials $K_n(x;p,N)$ \cite{krawt}, which obey the following recursion relation:
\begin{equation}
p(n-N)K_{n+1}(x;p,N)
-
\left[n(2p-1)+x-Np\right]K_{n}(x;p,N)
+
n(p-1)K_{n-1}(x;p,N)=0.
\label{kr}
\end{equation}
Comparing (\ref{kr}) with (\ref{rec2}) we see that
\begin{equation}
h^{(n)}_S(m)=\frac{2^{n/2}}{\sqrt{(2S-n)!}}K_n(m+S;1/2,2S).
\end{equation}

\newpage

\begin{thebibliography}{99}
\bibitem{kitagawa} M. Kitagawa and M. Ueda, Phys. Rev. A {\bf 47}, 5138 (1993).
\bibitem{walls}D. F. Walls and G. J. Milburn {\em Quantum Optics} (Spinger, Berlin, 1994).
\bibitem{phonons} J. Jansky and Y. Y. Yushin, Opt. Commun. {\bf 59}, 151 (1986); M. Artoni and J. L. Birman, Phys. Rev B. {\bf 44}, 3736 (1996); G. A. Garrett, A. G. Rojo, A. K. Sood, J. F. Whitaker and R. Merlin, Science {\bf 275}, 1628 (1997).
\bibitem{eberly} D. F. Walls and P. Zoller, Phys. Rev. Lett. {\bf 47}, 709 (1981). Wodkiewicz and J. Eberly, J. Opt. Soc. Am. B {\bf 2}, 458 (1985). 
\bibitem{wineland} D. J. Wineland, J. J. Bollinger, W. M. Itano and F. L. Moore, Phys. Rev. A {\bf 46}, R6797 (1992). 
\bibitem{raghavan} S. Raghavan, H. Pu, P. Meystre and N. P. Bigelow, Opt. Comm.{\bf 188}, 149 (2001).
\bibitem{yurke} This kind of states have been considered
 by B. Yurke [Phys. Rev. Lett. {\bf 56}, 1515 (1986)] for fermion interferometers 
and 
 by D. J. Wineland, J. J. Bollinger, W. M. Itano and D. J. Heinzen [Phys. Rev. A {\bf 50}, 56 (1994)] in the context of spin squeezing. 
\bibitem{law}  C. K. Law, H.T. Ng and P. T. Leung, Phys. Rev. A {\bf 63}, 055601 (2001).
\bibitem{milburn} G. J. Milburn, J. Corney, E. M. Wright and D. F. Walls, Phys. Rev. A {\bf 55}, 4318 (1997).
\bibitem{steel} M. J. Steel and M. J. Collett, Phys. Rev. A {\bf 57}, 2920 (1998).
\bibitem{gordon} D. Gordon and C. M. Savage,   Phys. Rev. A {\bf 59}, 4623 (1999).
\bibitem{feyn} R. P. Feynman, Phys. Rev. {\bf 56}, 340 (1939).
\bibitem{andre} A. Andr\'e and M. D. Lukin, Phys. Rev. A {\bf 65} 053819-1 (2002).
\bibitem{sorensen} A. S. S\o rensen and K. M\o lmer
Phys. Rev. Lett. {\bf 86}, 4431 (2001). 


\bibitem{oddn}
For odd $N$ the solutions are of the form
$\varphi_m\sim [{\Gamma(N)\over 
\Gamma(N/2-m) \Gamma(N/2+m)}]^{1/2} $.

\bibitem{arfken} G. B. Arfken and H. J. Weber, {\sl Mathematical Methods for Physicists}, Academic Press (1995), p.768.

\bibitem{kittel} C. Kittel, {\it Quantum Theory of Solids}, John Wiley \& Sons, (1987), p.50.
\bibitem{arecchi} F. T. Arecchi, E. Courtens, R. Gilmore and H. Thomas, Phys. Rev. A {\bf 6}, 2211 (1972) (see Appendix C).
\bibitem{rashid} M. A. Rashid, J. Math. Phys. {\bf 19}, 1391 (1978); {\bf 19}, 1397 (1978).
\bibitem{puri} G. S. Agrawal and R. R. Puri, Phys. Rev. A {\bf 49}, 4968 (1994).
\bibitem{krawt} See for example G. Szeg\"o, {\em Orthogonal Polynomials, 4th ed.} Providence, R.I.: Amer. Math. Soc., pp 35-37 (1975).
\end{thebibliography}
\end{document}